\documentclass[twocolumn,prd,showpacs,nofootinbib,a4]{revtex4}

\usepackage{graphicx}
\usepackage{epsfig}
\usepackage{textcomp}
\usepackage{color}

\def\theta{\vartheta}

\newcommand{\be}{\begin{equation}}
\newcommand{\ee}{\end{equation}}
\newcommand{\ba}{\begin{eqnarray}}
\newcommand{\ea}{\end{eqnarray}}
\newcommand{\lsim}   {\mathrel{\mathop{\kern 0pt \rlap
  {\raise.2ex\hbox{$<$}}}
  \lower.9ex\hbox{\kern-.190em $\sim$}}}
\newcommand{\gsim}   {\mathrel{\mathop{\kern 0pt \rlap
  {\raise.2ex\hbox{$>$}}}
  \lower.9ex\hbox{\kern-.190em $\sim$}}}

\begin{document}

\title{Stringent constraint on  neutrino Lorentz-invariance violation\\from the two IceCube PeV neutrinos}

\author{Enrico Borriello}
\affiliation{ II Institut f\"ur Theoretische Physik, Universit\"at Hamburg, Luruper Chaussee 149, 22761 Hamburg, Germany} 
\author{Sovan Chakraborty}
\affiliation{ II Institut f\"ur Theoretische Physik, Universit\"at Hamburg, Luruper Chaussee 149, 22761 Hamburg, Germany} 
\author{Alessandro Mirizzi}
\affiliation{ II Institut f\"ur Theoretische Physik, Universit\"at Hamburg, Luruper Chaussee 149, 22761 Hamburg, Germany} 
\author{Pasquale Dario Serpico} 
\affiliation{LAPTh, Univ. de Savoie, CNRS, B.P.110, Annecy-le-Vieux F-74941, France}

\begin{abstract}
It has been speculated that Lorentz-invariance violation (LIV) might be generated by quantum-gravity (QG) effects. As a consequence,  
particles  may not travel at the universal speed of light. In particular, superluminal extragalactic neutrinos would rapidly lose energy via the bremssthralung of electron-positron pairs $(\nu\to\nu\, e^+\, e^-)$, damping their initial energy into electromagnetic cascades, a figure constrained by Fermi-LAT data. We show that the two cascade  neutrino events with energies around 1~PeV  recently detected by IceCube---if attributed to extragalactic diffuse events, as it appears likely---can place \emph{ the strongest bound} on LIV in the neutrino sector, namely $\delta =(v^2-1) < {\cal O}(10^{-18})$, corresponding to a QG scale $M_{\rm QG} \gtrsim 10^{5} M_{\rm Pl}$ ($M_{\rm QG} \gtrsim 10^{-4} M_{\rm Pl}$) for a linear (quadratic) LIV, at least for models inducing superluminal neutrino effects ($\delta>0$).
 \end{abstract}

\pacs{  %03.30.+p 	%Special relativity
11.30.Cp, 	%Lorentz and Poincar� invariance
%14.60.Lm 	%Ordinary neutrinos
 95.85.Ry %Neutrino, muon, pion, and other elementary particles; cosmic rays
\hfill  LAPTH-015/13 }

\maketitle

%%%%%%%%%%%%%%%%%%%%%%%%%%%%%%%%%%%%%%%%%%%%%%%%%%%%%%%%%%%%%%%%%%%%%%%%
\section{Introduction}\label{Intro}
%%%%%%%%%%%%%%%%%%%%%%%%%%%%%%%%%%%%%%%%%%%%%%%%%%%%%%%%%%%%%%%%%%%%%%%%
It is conceivable that Lorentz-invariance might be violated in a candidate theory of  Quantum Gravity (QG), see  \cite{Liberati:2009pf} for a review.
In this context,  the space-time foam due to QG fluctuations might cause e.g. highly boosted energetic particles to propagate at speed $v$ different from the velocity of light $c$.
The resulting Lorentz-invariance violation (LIV) effect can be phenomenologically parametrized in terms of $\delta$ defined  as (we set $c=1$)
%.................................................................
\begin{equation}
\delta =\left(\frac{v}{v_0}\right)^2-1\,,\:\:\:v=\frac{\partial E}{\partial p}\,,v_0=\frac{p}{\sqrt{p^2+m^2}}\,,
\end{equation}
%................................................................. 
where, assuming that there is at least one  frame in which space and time
translations and spatial rotations are exact symmetries (typically the lab one),  one writes 
%.................................................................
\begin{equation}
E^2=p^2+m^2+f(p,\ldots)\,.
\end{equation}
%................................................................. 
The function $f$ is often expressed in powers of momentum over the phenomenological scale $M_{\rm QG}$ when LIV becomes
important, of the type $\propto p^\ell\,M_{\rm QG}^{2-\ell}$. Hence one can translate a constraint on $\delta$ into a constraint on $M_{\rm QG}$ via a relation of the kind
\begin{equation}
\delta =\left(\frac{v}{v_0}\right)^2-1\simeq \frac{v_0}{E}\frac{\partial f}{\partial p} \simeq \pm\left(\frac{E}{M_{\rm QG}}\right)^n\,\ ,\label{dispmod}
\end{equation}
with e.g.  cubic ($\ell=3$) or quartic ($\ell=4$) terms in powers of $p$ added to the dispersion relation of Eq.~(\ref{dispmod}), inducing ``linear'' ($n=1$) or ``quadratic'' ($n=2$) deviations from LI occurring at a mass scale $M_{\rm QG}$. The $+$ ($-$) implies superluminal (subluminal) velocity.

The observation of  LIV effects of this nature, if suppressed by the Planck scale $M_{\rm Pl} = 1.22\times 10^{28}$~eV, represents a very challenging task for any Earth-based experiment (see \cite{Kostelecky:2008ts} for a rewiev of the current bounds). In this respect, high-energy astrophysics experiments represent a promising, complementary tool able to probe the consequences of tiny LIV \cite{Mattingly:2005re,Liberati:2009pf}. Tests of LIV in photon propagation from objects like Gamma Ray Bursts (GRB) \cite{Ellis:2005}, Active Galactic Nuclei (AGN) \cite{Albert:2007qk} and Pulsars \cite{pulsar:1999} looking into delays in arrival times of photons of different energies can probe a quantum gravity mass scale up to  $M_{\rm QG}\sim 10^{26}$~eV for the linear case, and $M_{\rm QG}\sim 10^{19}$~eV for the quadratic case.

Experimental probes of LIV for neutrinos are limited by the scarcity of neutrino data from distant astrophysical sources.
In particular,  from SN1987A data, exploiting the flight delay  of the $\gamma$ with
respect to  $\nu$ of a few hours, it has   been obtained a bound of 
$\delta \lesssim 10^{-9}$ \cite{Stodolsky:1987vd}, corresponding to $M_{\rm QG} \gtrsim 10^{15}$~eV for $n=1$, or $M_{\rm QG} \gtrsim 10^{11}$~eV 
for $n=2$, taking $E\simeq 10$~MeV as typical SN $\nu$ energy. 
The supernova limit on LIV can reach $M_{\rm QG} \sim 10^{21}$ 
eV ($2 \times10^{14}$ eV) for the linear (quadratic) energy dependence, observing narrow time structure in the neutrino emission, like the ones associated 
with the neutronization burst~\cite{Chakraborty:2012gb}.
The bounds on LIV could be  further improved 
in case of detection of high-energy neutrinos ($E \gtrsim {\mathcal O} (\rm{GeV})$) from astrophysical 
sources, like  Gamma Ray Bursts or Active Galactic Nuclei~\cite{Jacob:2006gn}. Indeed, from the energy
dependent time of flight delay of neutrinos and the corresponding $\gamma$ rays, induced by LIV,
the limit that would be placed is $M_{QG} \gtrsim 10^{26}$~eV for the $n=1$ case, and 
$M_{QG} \gtrsim 10^{19}$~eV for the $n=2$ case, respectively.

For the specific case of  {\it superluminal} neutrinos, 
LIV would also allow processes that would be otherwise kinematically forbidden in
vacuum, namely neutrino Cherenkov radiation $(\nu \to \nu \, \gamma)$, neutrino splitting
$(\nu \to \nu\, \nu \,\bar\nu)$ and  bremssthralung of 
electron-positron pairs  $(\nu \to \nu \,e^+\, e^-)$ (see, e.g.~\cite{Maccione:2011fr}).  All these processes would produce
a depletion of the high-energy neutrino fluxes during their propagation. 
In particular, among the previous processes, the  neutrino pair production has been recognized as the fastest energy-loss process
for LIV neutrinos  and has been used in~\cite{Cohen:2011hx,GonzalezMestres:2011jc,Bi:2011nd} to invalidate the superluminal velocity claim made by the OPERA Collaboration~\cite{Adam:2011zb}. 
A similar analysis has also been able to put strong constraints on superluminal velocities of higher energy neutrinos.
The observation of  upward going atmospheric neutrino showers
(with path-length $L\simeq 500$ km), measured at $E\gtrsim 100$ TeV at the  IceCube experiment, has allowed to 
put the strongest bound on LIV, namely $\delta \lesssim 10^{-13}$~\cite{Cowsik:2012qm}.
That bound was derived by comparing the observed spectrum with the theoretical predicted one. It was noted before that a conceptually similar
method could lead to improvements if it could be applied to the spectrum of an astrophysical source such as a supernova remnant, as noted in~\cite{Bi:2011nd}.
In Sec.~II, following from recent {\it observations} of IceCube, we explore a new stringent bound based on this particle physics mechanism but following from a different,
{\it calorimetric} (as opposed to {\it spectral}) diagnostic tool, based on multi-messenger considerations involving diffuse gamma-ray data.
In Sec.~III we discuss these results and conclude.

%%%%%%%%%%%%%%%%%%%%%%%%%%%%%%%%%%%%%%%%%%%%%%%%%%%%%%%%%%%%%%%%%%%%%%%%
\section{New stringent bound}\label{bound}
%%%%%%%%%%%%%%%%%%%%%%%%%%%%%%%%%%%%%%%%%%%%%%%%%%%%%%%%%%%%%%%%%%%%%%%%

Since the effects of LIV processes  typically grow with energy and distance, 
it is worthwhile to exploit high-energy neutrinos traveling along  a sufficiently long baseline.
It has been argued for example that observations of the cosmogenic neutrino background (produced by ultra-high energy
cosmic ray losses) might lead to stringent constraints, see~\cite{Mattingly:2009jf}.
In this context, the IceCube experiment has recently reported the detection of two cascade neutrino events with energies
around $E \simeq 1$ PeV in the period of data taking 2011-2012~\cite{Aartsen:2013bka}. 
At the moment, the origin of these events is not settled. It seems that conventional atmospheric neutrinos
or cosmogenic neutrinos~\cite{Roulet:2012rv} are unlikely to be the sources of these events, and that the chance that these events 
are associated to prompt atmospheric neutrinos from heavy quark decays is at most a few \%~\cite{Cholis:2012kq}.  
There is also another intriguing experimental indication: an IceCube analysis of neutrino-induced muon track events  shows that, 
although the sample is  dominated by  conventional atmospheric neutrinos,  data do prefer marginally (at the 1.8 $\sigma$ level) an extra component at $E> 100\,$TeV~\cite{ASnow2012}. Albeit of little significance at present, it is intriguing that the extrapolation of this best-fit flux to PeV scale is in perfect agreement with the observed two PeV cascade events mentioned above. Although the significance of an additional component is still low, the hypothesis that these two neutrinos could be the first indication of a diffuse extragalactic astrophysical flux at the PeV scale appears at the same time the most likely and the most exciting one\footnote{Although the referred analysis is based on the more advanced stage/statistics of IceCube-59, it is worth reminding that different analysis of IceCube-40 public data show contradicting results. In particular, an excess can be seen in $\nu_e$ data \cite{Abbasi:2011jx}, but not in $\nu_{\mu}$ data \cite{Abbasi:2011zx}.}.

In this section, we deduce an important  improvement over the current bounds on superluminal neutrino velocity, following from the above-mentioned assumption. In Sec.~\ref{concl} we shall comment further on this point. 

First, let us remind that for $\delta>0$ the process $\nu \to \nu\, e^+\, e^-$ is kinematically allowed  provided that 
\begin{equation}
E_\nu\gtrsim 2\,m_e/\sqrt{\delta}\simeq {\rm PeV}\sqrt{10^{-18}/\delta}\label{condition}
\end{equation} 
and assuming LI conservation in the electron sector~\cite{Cohen:2011hx}. This process implies a  pair
production decay rate of~\cite{Cohen:2011hx}: 
\begin{equation}
\Gamma_{e^\pm} =\frac{1}{14}\frac{G_F^2 E^5 \delta^3}{192\,\pi^3}=2.55\times 10^{53}\delta^3 E_{\rm PeV}^5\,\,{\rm Mpc}^{-1}.\label{mainproc}
\end{equation} 
As in~\cite{Cohen:2011hx}, we neglect here the process $(\nu \to \nu\, \nu\, \bar\nu)$, which is only expected to bring minor modification to our argument.
On the other hand, if the process associated with the rate of Eq.~(\ref{mainproc}) is forbidden because of threshold effects, the $\nu \to \nu \gamma$ is anyway operational 
and a channel for energy losses, although it is two to three orders of magnitude less efficient than $\nu \to \nu\, e^+ e^-$.

Remarkably, without any significant additional information on astrophysical sources of neutrinos (neither on the specific mechanism of neutrino production), this is sufficient to derive a strong bound. Let us assume that the
PeV neutrinos are generated somehow in extragalactic sources. If the above processes are effective, the neutrino flux will be soon depleted at the expense
of injecting electromagnetic energy while propagating to the Earth. The interaction length of $e^{\pm}$ onto CMB photons is extremely short, of the order of few kpc.
Even a gamma ray will not propagate much more than $\mathcal{O}$(10) kpc at these energies before pair-producing on the CMB (see for example~\cite{Lee:1996fp}).
The inverse-Compton scattering photons induced by  $e^{\pm}$ will populate a gamma-ray flux at whose energy is roughly stored between $E_1\sim {\cal O}(1)$ GeV and $E_2 {\cal O}(100)$ GeV,
see for example Fig. 5 in~\cite{Semikoz:2003wv}. 
This flux is constrained by Fermi
data~\cite{Abdo:2010nz} to have an {\it integrated} energy density 
\begin{equation}
\omega_\gamma\,=\,\frac{4\pi}{c}\hspace{-6pt}\int_{E_1}^{E_2}
 E\, \frac{d \varphi_{\gamma}}{dE}\, \mathrm{d}E\, \lesssim\, 5.7\times 10^{-7} \,\textrm{eV/cm}^{3} \, . 
\end{equation}
We are taking the normalization for the extragalactic flux from Fermi and the spectral shape for the energy density $d \varphi_{\gamma}/dE\sim E^{-2.41}$, just like in previous
publications on this subject such as \cite{Berezinsky:2010xa}. Note also that the unaccounted flux (not reasonably attributed to unresolved astrophysical sources) is probably only a fraction of the above one, see e.g.~\cite{Calore:2013yia} for a recent modeling of that.
Now, the two events detected in IceCube in two years of data taking would imply a diffuse energy flux $E^{2}_{\nu}\, d \varphi_E/dE \simeq 3.6 \times 10^{-8}$ GeV cm$^{-2}$ s$^{-1}$ sr$^{-1}$~\cite{Aartsen:2013bka}, actually quite close to the so-called Waxman-Bahcall benchmark~\cite{Waxman:1998yy}.
These numbers imply that one cannot tolerate an energy density $\omega_\nu^{\rm in}$ in the {\it initial} neutrino flux larger by about two orders of magnitude than the observed one 
\begin{equation}
\omega_{\nu}^{\rm obs}\,=\,\frac{4\pi}{c}\hspace{-6pt}\int\limits_{1\,\textrm{\scriptsize{PeV}}}^{1.2\,\textrm{\scriptsize{PeV}}}\hspace{-6pt} 
 E\, \frac{d \varphi_{E}}{dE}\, \mathrm{d}E\, \simeq\, 2.7\times 10^{-9} \,\textrm{eV/cm}^{3} \, ,
\end{equation}
otherwise the electromagnetic energy injected  via the process would basically saturate the bound. Note that we did not extrapolate the observed flux beyond the energy window at which
the two events have been measured. Once accounting for the fact that the process in question transfers a large fraction of the initial neutrino energy into the $e^\pm$ pair~\cite{Cohen:2011hx},  a simple back-of-the envelope calculation leads to an approximate constraint of the type:
\begin{equation}
e^{-\Gamma\,d} \gtrsim \frac{\omega_\nu^{\rm obs}}{\omega_\gamma}\sim10^{-2}\,,
\end{equation} 
where $d$ is the characteristic distance of the sources. Provided that the channel  $\nu \to \nu\, e^+ e^-$ is open, we get 
\begin{equation}
\delta_{e^{\pm}}^3\,d_{\rm Mpc}< 1.8\times 10^{-53}\,.\label{bound}
%\delta_{\gamma}^3\,d_{\rm Mpc}< 0.8\times 10^{-50}\,.
\end{equation} 
For the simplest scenario of cosmologically distant sources, a reasonable value is $d\sim {\cal O}(10^3)\,$Mpc. 
In fact the emissivity of a diffuse neutrino flux from cosmologically distributed sources (such as GRBs) is peaked at redshift $z\sim1$ \cite{Iocco:2007td}, that corresponds to a comoving distance of more than 3 Gpc. 
Equation~(\ref{bound}) nominally implies $\delta \lesssim {\cal O}(10^{-19})$, which means that the pair-production mechanism is not operational and that the actual bound is thus  $\delta <{\cal O}(10^{-18})$ (see the condition of Eq.~(\ref{condition})). A priori, the $\nu \to \nu \gamma$ is kinematically accessible to PeV neutrinos even for $\delta <{\cal O}(10^{-18})$ and depending on its rate it could put a more stringent bound, but it is easy to check that  it is not the case: it leads instead to a slightly weaker (albeit of comparable order of magnitude) bound.
However, the latter channel allows  to infer that the derived bound probably holds under more general hypotheses than the ones we assumed: for example it should not depend crucially
on the Ansatz that the LIV in the electron sector is much smaller than in the neutrino one.

%%%%%%%%%%%%%%%%%%%%%%%%%%%%%%%%%%%%%%%%%%%%%%%%%%%%%%%%%%%%%%%%%%%%%%%%
\section{Discussion and conclusion}\label{concl}
%%%%%%%%%%%%%%%%%%%%%%%%%%%%%%%%%%%%%%%%%%%%%%%%%%%%%%%%%%%%%%%%%%%%%%%%
We have derived a very stringent bound on LIV in the neutrino sector, $\delta<{\cal O}(10^{-18})$, from the observations of two PeV scale IceCube events and remarkably few assumptions
on the underlying astrophysical sources. 

It is important to emphasize that our argument is neither based on a study of the neutrino flux shape distortion, like in \cite{Cowsik:2012qm} nor on the assumption of any putative Galactic source, like in \cite{Bi:2011nd}, as none is currently plausible to explain the observed flux we refer to. What we propose here is instead a novel ``calorimetric'' bound, based on the observed extragalactic gamma ray diffuse flux. 

Let us now discuss the model dependence in more detail. 
We just assumed that the events are due to an extragalactic flux, hence with characteristic distance of source(s) of the order of a Gpc.  Of course, once additional
information will be available (e.g. on the number density and redshift distribution of the sources contributing to the flux) an improved calculation will be possible. 
We do not expect however that the ignorance of these details affects the bound very much. In fact, within our approximations the bound on $\delta$ scales as
\begin{equation}
\delta\lesssim 10^{-18}\times{\rm Max}\left[ 1.6 \left(\frac{\ln(\omega_\gamma/\omega_\nu^{\rm obs})}{E_{\rm PeV}^5\,d_{\rm Mpc}}\right)^{\frac{1}{3}},\,E_{\rm PeV}^{-2}\right]\, ,
\end{equation} 
with a very weak dependence on the initial assumption on the fluxes, as well as a dependence only on the cubic root of the typical distance.
For example, it is worth noting that even in the extreme (and unlikely) case where the observed neutrino flux were dominated by the closest extragalactic
source candidates, at distances of the order of several Mpc---Centaurus A, which is the closest one, at about 4 Mpc from us---one would still deduce $\delta\lesssim {\cal O}(10^{-18})$. Again, this is basically suggesting that the energy loss channel via pair emission must be a closed channel. On the other hand, detecting higher energy events would improve and make the bounds more robust.

The extragalactic nature is essential in making the electromagnetic ``cascade'' argument operational, but actually it is also the most likely explanation,  given that the flux level required
to match observations appear close to predicted diffuse fluxes (see e.g.~\cite{Becker:2007sv}).
Not only that:  an indirect argument against a closer astrophysical origin is that the Galactic diffuse flux at PeV energy can be computed quite reliably and is expected to be much smaller, see e.g.~\cite{Evoli:2007iy}. This is true provided that the source is relatively ``thin'' to cosmic rays, so that the flux is dominated by cosmic ray production in interstellar medium.
This appears almost unavoidable, especially since we would need accelerators capable of maximal energies up to 10$^{16}$ eV/nucleon at least, which would be hard to impossible
to achieve with large energy losses within the accelerator. Anyway,  some bound should exist also if one or more hypothetical Galactic sources contribute to this flux. Lacking promising candidates of this type in the literature, we do not consider this further. In general, the bound should be then obtained by estimating an input flux upper bound by some other sort 
of multi-messenger constraint: Typically,  other by-products (or tertiaries of their energy degradation) of the neutrino production processes at the source.

Finally, what if the events should be eventually attributed to atmospheric background? We briefly mentioned that this appears unlikely at present. Yet, it is worth noting that even in that case, the current bound~\cite{Cowsik:2012qm} could be still improved by a more modest amount (perhaps one order of magnitude), although a more detailed analysis of the type of~\cite{Cowsik:2012qm}---properly accounting for prompt contribution and its uncertainty---would be needed to draw a more quantitative conclusion.

In summary, we have argued that a confirmation of the extragalactic astrophysical nature of the PeV events detected by IceCube would not only open a new window to the high-energy universe, but also allow a significant jump in tests of fundamental physics. To obtain this result, as more and more customary in astroparticle physics, we relied on a multi-messenger strategy (here the link with diffuse gamma ray background), which proves once again the power of this approach.

%%%%%%%%%%%%%%%%%%%%%%%%%%%%%%%%%%%%%%%%%%%%%%%%%%%%%%%%%%%%%%%%%%%%%%%%
%% Acknowledgments %%%%%%%%%%%%%%%%%%%%%%%%%%%%%%%%%%%%%%%%%%%%%%%%%%%%%
%%%%%%%%%%%%%%%%%%%%%%%%%%%%%%%%%%%%%%%%%%%%%%%%%%%%%%%%%%%%%%%%%%%%%%%%
\begin{acknowledgments}
The authors thank Basudeb Dasgupta, Tom Gaisser, Teresa Montaruli, and Anne Schukraft for useful conversations, Jorge S. Diaz for feedback on the manuscript, and particularly
Luca Maccione for comments. The work of E.B. and S.C. and A.M.  was supported by the German Science Foundation (DFG) within the Collaborative Research Center 676 ``Particles, Strings and the Early Universe''. At LAPTh, this activity was developed coherently with the research axes supported by the Labex grant ENIGMASS.

\end{acknowledgments}

%%%%%%%%%%%%%%%%%%%%%%%%%%%%%%%%%%%%%%%%%%%%%%%%%%%%%%%%%%%%%%%%%%%%%%%%
%% References %%%%%%%%%%%%%%%%%%%%%%%%%%%%%%%%%%%%%%%%%%%%%%%%%%%%%%%%%%
%%%%%%%%%%%%%%%%%%%%%%%%%%%%%%%%%%%%%%%%%%%%%%%%%%%%%%%%%%%%%%%%%%%%%%%%


\begin{thebibliography}{99}


\bibitem{Liberati:2009pf}
S.~Liberati and L.~Maccione, 
% {\it {Lorentz Violation: Motivation and New   Constraints}},  
Annu. Rev. Nucl. Part. Sci. {\bf 59} (2009) 245--267.

%\cite{Kostelecky:2008ts}
\bibitem{Kostelecky:2008ts}
  V.~A.~Kostelecky and N.~Russell,
  %``Data Tables for Lorentz and CPT Violation,''
  Rev.\ Mod.\ Phys.\  {\bf 83} (2011) 11.                                
  %[arXiv:0801.0287 [hep-ph]].
  %%CITATION = ARXIV:0801.0287;%%
  %267 citations counted in INSPIRE as of 07 May 2013


%\cite{Mattingly:2005re}
\bibitem{Mattingly:2005re} 
  D.~Mattingly,
  %``Modern tests of Lorentz invariance,'' 
 Living Rev.\ Rel.\  {\bf 8}, 5 (2005).  %%CITATION = GR-QC/0502097;%%

\bibitem{Ellis:2005}
J.~R.~Ellis, N.~E.~Mavromatos, D.~V.~Nanopoulos,
A.~S.~Sakharov and E.~K.~G.~Sarkisyan,
  %``Robust Limits on Lorentz Violation from Gamma-Ray Bursts,''
  Astropart.\ Phys.\  {\bf 25},  402 (2006)
  [Astropart.\ Phys.\  {\bf 29}, 158 (2008)]
% ; ibid J.~Ellis, N.~E.~Mavromatos,
%   D.~V.~Nanopoulos, A.~S.~Sakharov and E.~K.~G.~Sarkisyan.
  %``Erratum (astro-ph/0510172): Robust Limits on Lorentz Violation from
  %Gamma-Ray Bursts,''

%\cite{Albert:2007qk}
\bibitem{Albert:2007qk} 
  J.~Albert {\it et al.}  [MAGIC and Other Contributors Collaborations],
  %``Probing Quantum Gravity using Photons from a flare of the active galactic nucleus Markarian 501 Observed by the MAGIC telescope,''  
Phys.\ Lett.\ B {\bf 668}, 253 (2008).
  %%CITATION = ARXIV:0708.2889;%%

%\cite{Kaaret:1999ve}
\bibitem{pulsar:1999} 
  P.~Kaaret,
  %``Pulsar radiation and quantum gravity,'' 
 astro-ph/9903464.  
%%CITATION = ASTRO-PH/9903464;%%



%\cite{Stodolsky:1987vd}
\bibitem{Stodolsky:1987vd} 
  L.~Stodolsky,
  %``The Speed Of Light And The Speed Of Neutrinos,''  
Phys.\ Lett.\ B {\bf 201}, 353 (1988).  %%CITATION = PHLTA,B201,353;%%



%\cite{Chakraborty:2012gb}
\bibitem{Chakraborty:2012gb} 
  S.~Chakraborty, A.~Mirizzi and G.~Sigl,
 % ``Testing Lorentz Invariance with neutrino burst from supernova neutronization,'' 
 Phys.\ Rev.\ D {\bf 87}, 017302 (2013).  %%CITATION = ARXIV:1211.7069;%%



%\JacobGN
\bibitem{Jacob:2006gn}
  U.~Jacob and T.~Piran,
  %``GRBs neutrinos as a tool to explore quantum gravity induced Lorentz violation,''
Nature Phys.\  {\bf 3}, 87 (2007).
%%CITATION = hep-ph/0607145%%

%\cite{Maccione:2011fr}
\bibitem{Maccione:2011fr}
  L.~Maccione, S.~Liberati and D.~M.~Mattingly,
  %``Violations of Lorentz invariance in the neutrino sector after OPERA,''
  arXiv:1110.0783 [hep-ph].
  %%CITATION = ARXIV:1110.0783;%%
  %28 citations counted in INSPIRE as of 21 Mar 2013


%\cite{Cohen:2011hx}
\bibitem{Cohen:2011hx} 
  A.~G.~Cohen and S.~L.~Glashow,
%  ``Pair Creation Constrains Superluminal Neutrino Propagation,''  
Phys.\ Rev.\ Lett.\  {\bf 107}, 181803 (2011).  
%[arXiv:1109.6562 [hep-ph]].  %%CITATION = ARXIV:1109.6562;%%

  %\cite{GonzalezMestres:2011jc}
\bibitem{GonzalezMestres:2011jc} 
  L.~Gonzalez-Mestres,
  %``Astrophysical consequences of the OPERA superluminal neutrino,''
  arXiv:1109.6630 [physics.gen-ph].
  %%CITATION = ARXIV:1109.6630;%%

%\cite{Bi:2011nd}
\bibitem{Bi:2011nd}
  X.~-J.~Bi, P.~-F.~Yin, Z.~-H.~Yu and Q.~Yuan,
  %``Constraints and tests of the OPERA superluminal neutrinos,''
  Phys.\ Rev.\ Lett.\  {\bf 107} (2011) 241802.                             
%  [arXiv:1109.6667 [hep-ph]] 
  %%CITATION = ARXIV:1109.6667;%%




%\cite{Adam:2011zb}
\bibitem{Adam:2011zb}
  T.~Adam {\it et al.}  [OPERA Collaboration],
  %``Measurement of the neutrino velocity with the OPERA detector in the CNGS beam,''
  JHEP {\bf 1210} (2012) 093.
  %[arXiv:1109.4897 [hep-ex]].
  %%CITATION = ARXIV:1109.4897;%%
  %306 citations counted in INSPIRE as of 21 Mar 2013

%\cite{Cowsik:2012qm}
\bibitem{Cowsik:2012qm} 
  R.~Cowsik, T.~Madziwa-Nussinov, S.~Nussinov and U.~Sarkar,
  %``Testing Violations of Lorentz Invariance with Cosmic-Rays,'' 
 Phys.\ Rev.\ D {\bf 86}, 045024 (2012).  
%[arXiv:1206.0713 [hep-ph]].  %%CITATION = ARXIV:1206.0713;%%




\bibitem{Mattingly:2009jf} 
  D.~M.~Mattingly, L.~Maccione, M.~Galaverni, S.~Liberati and G.~Sigl,
  %``Possible cosmogenic neutrino constraints on Planck-scale Lorentz violation,''
  JCAP {\bf 1002}, 007 (2010).
  %[arXiv:0911.0521 [hep-ph]].
  %%CITATION = ARXIV:0911.0521;%%
  
%\cite{Aartsen:2013bka}
\bibitem{Aartsen:2013bka}
  M.~G.~Aartsen {\it et al.}  [IceCube Collaboration],
  %``First observation of PeV-energy neutrinos with IceCube,''
  arXiv:1304.5356 [astro-ph.HE].                                         
  %%CITATION = ARXIV:1304.5356;%%



%\cite{Roulet:2012rv}
\bibitem{Roulet:2012rv} 
  E.~Roulet, G.~Sigl, A.~van Vliet and S.~Mollerach,
  %``PeV neutrinos from the propagation of ultra-high energy cosmic rays,''  
JCAP {\bf 1301}, 028 (2013). 
% [arXiv:1209.4033 [astro-ph.HE]].  %%CITATION = ARXIV:1209.4033;%%

%\cite{Cholis:2012kq}
\bibitem{Cholis:2012kq} 
  I.~Cholis and D.~Hooper,
  %``On The Origin of IceCube's PeV Neutrinos,'' 
 arXiv:1211.1974 [astro-ph.HE].  %%CITATION = ARXIV:1211.1974;%%


\bibitem{ASnow2012} Anne Schukraft, talk at NOW 2012, to appear  Nucl. Phys. B Proc. Supplements. Slides available at \texttt{www.ba.infn.it/~now/ }



%\cite{Abbasi:2011jx}
\bibitem{Abbasi:2011jx}
  R.~Abbasi {\it et al.}  [IceCube Collaboration],
  %``A Search for a Diffuse Flux of Astrophysical Muon Neutrinos with the IceCube 40-String Detector,''
  Phys.\ Rev.\ D {\bf 84} (2011) 082001.                          
  %[arXiv:1104.5187 [astro-ph.HE]].               
  %%CITATION = ARXIV:1104.5187;%%
  %51 citations counted in INSPIRE as of 07 May 2013


%\cite{Abbasi:2011zx}
\bibitem{Abbasi:2011zx}
  R.~Abbasi {\it et al.}  [IceCube Collaboration],
  %``The IceCube Neutrino Observatory II: All Sky Searches: Atmospheric, Diffuse and EHE,''
  arXiv:1111.2736 [astro-ph.HE].                       
  %%CITATION = ARXIV:1111.2736;%%
  %9 citations counted in INSPIRE as of 07 May 2013



   \bibitem{Lee:1996fp} 
  S.~Lee,
  %``On the propagation of extragalactic high-energy cosmic and gamma-rays,''
  Phys.\ Rev.\ D {\bf 58}, 043004 (1998).
  %[astro-ph/9604098].

\bibitem{Semikoz:2003wv} 
  D.~V.~Semikoz and G.~Sigl,
  %``Ultrahigh-energy neutrino fluxes: New constraints and implications,''
  JCAP {\bf 0404}, 003 (2004)
  [hep-ph/0309328].
  %%CITATION = HEP-PH/0309328;%%
  
  
\bibitem{Abdo:2010nz} 
  A.~A.~Abdo {\it et al.}  [Fermi-LAT Collaboration],
  %``The Spectrum of the Isotropic Diffuse Gamma-Ray Emission Derived From First-Year Fermi Large Area Telescope Data,''
  Phys.\ Rev.\ Lett.\  {\bf 104}, 101101 (2010)
  %[arXiv:1002.3603 [astro-ph.HE]].
  %%CITATION = ARXIV:1002.3603;%%

\bibitem{Berezinsky:2010xa} 
  V.~Berezinsky, A.~Gazizov, M.~Kachelriess and S.~Ostapchenko,
  %``Restricting UHECRs and cosmogenic neutrinos with Fermi-LAT,''
  Phys.\ Lett.\ B {\bf 695}, 13 (2011).
  %[arXiv:1003.1496 [astro-ph.HE]].
  %%CITATION = ARXIV:1003.1496;%%
  
  \bibitem{Calore:2013yia} 
  F.~Calore, M.~Di Mauro and F.~Donato,
  %``Updated constraints on WIMP dark matter annihilation into gamma-rays,''
  arXiv:1303.3284 [astro-ph.CO].
  %%CITATION = ARXIV:1303.3284;%%
 
 %\cite{Waxman:1998yy}
\bibitem{Waxman:1998yy} 
  E.~Waxman and J.~N.~Bahcall,
  %``High-energy neutrinos from astrophysical sources: An Upper bound,'' 
 Phys.\ Rev.\ D {\bf 59}, 023002 (1999).
%  [hep-ph/9807282].  %%CITATION = HEP-PH/9807282;%%


%\cite{Iocco:2007td}
\bibitem{Iocco:2007td}
  F.~Iocco, K.~Murase, S.~Nagataki and P.~D.~Serpico,
  %``High Energy neutrino signals from the Epoch of Reionization,''
  Astrophys.\ J.\  {\bf 675} (2008) 937.                                      
%  [arXiv:0707.0515 [astro-ph]].
  %%CITATION = ARXIV:0707.0515;%%
  %6 citations counted in INSPIRE as of 06 May 2013


\bibitem{Becker:2007sv} 
  J.~K.~Becker,
  %``High-energy neutrinos in the context of multimessenger physics,''
  Phys.\ Rept.\  {\bf 458}, 173 (2008).
  %[arXiv:0710.1557 [astro-ph]].
  %%CITATION = ARXIV:0710.1557;%%

\bibitem{Evoli:2007iy} 
  C.~Evoli, D.~Grasso and L.~Maccione,
  %``Diffuse Neutrino and Gamma-ray Emissions of the Galaxy above the TeV,''
  JCAP {\bf 0706}, 003 (2007).
  %[astro-ph/0701856].
  %%CITATION = ASTRO-PH/0701856;%%



\end{thebibliography}
\end{document}